\documentclass[aps,prb,twocolumn,longbibliography]{revtex4-1}
\usepackage{epsfig,amsopn}
\usepackage{graphicx}
\usepackage{lipsum}
\usepackage{color}
\usepackage{amsmath}
\usepackage{amssymb}
\usepackage{enumerate}
\newcommand\bea{\begin{eqnarray}}
\newcommand\eea{\end{eqnarray}}
\newcommand\beq{\begin{equation}}
\newcommand\eeq{\end{equation}}

\def\nn{\nonumber}
\def\f{\frac}

\def\om{\omega}

\def\ga{\gamma}

\def\De{\Delta}
\def\dg{\dagger}
\def\la{\langle}
\def\ra{\rangle}

\begin{document}
\title{Anomalous Josephson effect and rectification in junctions between   Floquet topological superconductors} 
\author{ Abhiram Soori~~}  
\email{abhirams@uohyd.ac.in}
\affiliation{ School of Physics, University of Hyderabad, Prof. C. R. Rao Road, 
Gachibowli, Hyderabad-500046, India.}
\begin{abstract}
Periodically driven Kitaev chains are  known to exhibit novel Floquet Majorana fermions and anomalous Floquet end modes. 
The fact that `quasienergy in Floquet systems is periodic' poses a difficulty in defining the ground state in periodically driven systems. To overcome this problem, we start with the ground state of the undriven Kitaev chain and gradually switch on the periodic driving in chemical potential over a timescale $\tau$. The extent to which the single particle eigenstates of the undriven system get distributed among the Floquet states upon driving can be characterized by inverse participation ratio. In a Josephson junction between two periodically driven Kitaev chains not differing in phase of the pair potentials but differing in  phases of the driving potentials, a net average current flows from one superconductor to the other. We term such a current anomalous current. Further, we study current phase relation in junctions between two periodically driven superconductors and find that the system exhibits nonequilibrium Josephson diode effect. The Floquet Majorana end modes and anomalous Floquet end modes whenever present contribute significantly to the anomalous current and the diode effect. Further, the anomalous current and the nonequilibrium Josephson diode effect survive when the periodic driving is adiabatically switched on.  
\end{abstract}
\maketitle

\section{Introduction}
Floquet topological matter has drawn the attention of researchers in the last decade since novel phases of matter can be realized in these systems~\cite{lindner2011,jiang11,thakurathi13,benito14,oka19,liu19,harper20,ghosh21,bandho21,agarwal22,peng21,mondal22}. Physical properties of a system change with time when the  Hamiltonian is time dependent. However, for a system whose Hamiltonian changes with time periodically, the physical quantities of interest when averaged over sufficiently long time are determined by eigenstates of the unitary time evolution operator $U_{T}$ that takes the state of the system from time $t=0$ to $t=T$, where $T$ is the period. Such systems are termed Floquet systems, and the eigenstates of $U_T$ are called Floquet states.
A Kitaev chain driven periodically exhibits a novel topological end mode termed $\pi$-Floquet Majorana fermion, which is an eigenstate of $U_T$ with an eigenvalue $-1$. Such a topological mode has no counterpart in the undriven system. Periodically driving planar Josephson junctions is believed to generate Floquet Majorana fermions~\cite{liu19}. Also, biased planar Josephson junction can generate Floquet Majorana fermions due to AC Josephson effect~\cite{peng21}.   Anomalous end modes that are not topological can also exist in certain periodically driven Kitaev chains~\cite{saha17}.
Adiabatically turning on the driving in Floquet systems is known to produce  states that are similar in spirit to ground states of equilibrium systems~\cite{kitagawa11,heinisch16,bandho19,bandho21}.   

On average, a net current can be driven by periodic in time potentials applied to small regions in transport channel connected to reservoirs~\cite{thouless83,brouwer98,switkes99,brouwer01,avron01,moskalets02,agarwal07,agarwal07prb,soori10}. This phenomenon known as quantum charge pumping has also been extended to superconducting systems~\cite{blaauboer02,governale05,soori20pump}. A recent surge of activities in superconducting diode effect~\cite{waka17,qin17,siva18,lust18,hoshino18,yasuda19,ando2020,ita2020,baum2022,moodera,wu2022,souto} was followed by  investigation of Josephson diode effect in periodically driven Josephson junctions~\cite{soori22sde,paaske22}. The maximum and minimum values of Josephson current in the current phase relation of a Josephson junction not being equal in magnitude and opposite in sign marks Josephson diode effect. When the junction between two superconductors is periodically driven, the current varies as a function of time and the current averaged over (infinitely) long time quantifies the charge transferred from one superconductor to the other. Such a long time averaged current is essentially carried by Floquet states. 
Josephson junctions consisting of regions with spin orbit coupling and Zeeman field are known to exhibit a nonzero Josephson current in the absence of a superconducting phase difference - an effect known as anomalous Josephson effect~\cite{yokoyama14,campagnano15,mintillo18}. These developments motivate us to study anomalous current and nonequilibrium Josephson diode effect in junctions between Floquet superconductors. Periodically driven Kitaev chain offers a rich playground since it hosts different phases wherein there can be no Floquet Majorana fermions or multiple pairs of Floquet Majorana fermions or also anomalous Floquet end modes depending on the choice of parameters. The role played by Floquet Majorana fermions and anomalous end modes is an interesting aspect which can be studied in periodically driven Kitaev chain. 

In this work, we study the Kitaev chains, wherein the chemical potential changes periodically in time. First, we study topology and end modes of a periodically driven Kitaev chain. We numerically calculate the winding number and study the end modes of an open chain. We find Floquet end modes and anomalous non-topological end modes. Then, we study how the equilibrium ground state of the undriven chain evolves into nonequilibrium state of a periodically driven system when the periodic driving is switched on gradually over a timescale. We characterize the nonequilibrium state after the driving is switched on completely by calculating  inverse participation ratio. We then study the long time averaged current at the junction between two Floquet superconductors in the two limits: when the driving is switched on suddenly and when the driving is switched on gradually over a timescale. We find a novel effect akin to Josephson effect wherein a net long time averaged current flows between two Floquet superconductors that are driven with a difference in phase of the driving potential even when there is no difference in the phases of the superconducting pair potential. We calculate a weighted current that draws significantly high contribution from the Floquet Majorana end modes to quantify the current carried by the Majorana modes.  We also study the current phase relation and investigate diode effect. We study these effects as a function of the timescale over which the driving is switched on. 

  \section{Model and calculation}~\label{sec-model}
  We first study a single periodically driven Kitaev chain, followed by calculations on a junction between two driven Kitaev chains. The Hamiltonian for a driven Kitaev chain can be written as 
  \bea
  H_K(t) &=& H_{K,0}+H_{K,1}(t), {\rm ~for~} t\ge 0, \nn \\
  H_{K,0} &=& -w_h\sum_{n=1}^{L_S-1}(c^{\dg}_{n+1}\tau_zc_{n}+{\rm h.c.})
 -\mu\sum_{n=1}^{L_S}c^{\dg}_{n} \tau_z c_{n}\nn\\
 &&+\De \sum_{n=1}^{L_S-1}(c^{\dg}_{n+1}\tau_x c_{n}+{\rm h.c.}), ~~~\nn \\
 H_{K,1}(t) &=& V_0 \cos{\om t} \sum_{n=1}^{L_S}c^{\dg}_{n}\tau_zc_{n}, \nn \\ \label{eq:H-KC}
  \eea
  where $w_h$ is the hopping amplitude, $\mu$ is the chemical potential, $\De$ is the strength of $p$-wave pairing, $V_0$ is the amplitude of the driving potential, $\om$ is the frequency of driving, $L_S$ is the number of sites in the Kitaev chain,  $c_n=[d_n,d^{\dg}_n]^T$, $d_n$ is the annihilation operator for an electron at site $n$ and $\tau_j$ ($j=x, y, z$) are Pauli spin matrices that act on the particle hole space. The subscript $K$ is used to specify that the Hamiltonian in Eq.~\eqref{eq:H-KC} describes a single Kitaev chain. 
 A state of the system evolves from time $t_1$ to $t_2$ under the unitary time evolution operator $U_K(t_2,t_1)$ determined by the Hamiltonian of the system. The eigenstates of $U_K(T,0)$  ($T=2\pi/\om$ is the period) are called the Floquet  eigenstates and the eigenvalues of $U_{K}(T,0)$ are called Floquet eigenvalues.

\begin{figure}
  \includegraphics[width=8cm]{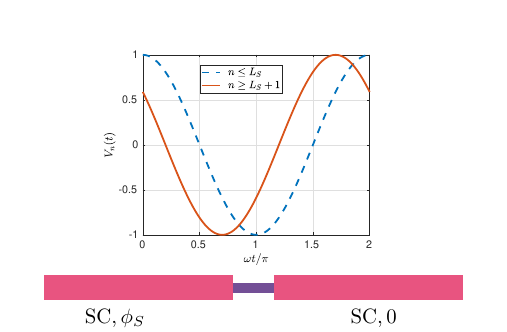}
  \caption{Schematic of the junction between two Floquet superconductors described by Kitaev model. Each Kitaev chain is a lattice with $L_S$ sites. The onsite potential at site $n$ is $\mu-V_n(t)$. In addition to a superconducting phase difference $\phi_S$ between the Kitaev chains, the chemical potentials of the two chains are driven with a difference $\phi$ in their phase.}\label{fig:schem}
 \end{figure}

Josephson junctions are typically described by a model that connects two semi-infinite superconductors. In such junctions, bound states localized at the junction within the superconducting gap carry Josephson current in contrast to the bound states in normal metals which carry no current~\cite{furusaki99}. However, the Andreev bound states decay exponentially in the superconductor away from the junction and  superconductors of finite length larger than the superconducting coherence length mimics the junction between semi-infinite superconductors. However, for a junction between finite superconductors, current contributions from all states need to be accounted, though the subgap state localized at the junction contributes significantly. In this work, we consider a junction between superconductors of finite length.  The Hamiltonian for a junction between two Kitaev chains depicted in Fig.~\ref{fig:schem} for time $t>0$ is given by 
\bea  H(t) &=& H_{0}+H_{1}(t),~\nn \\
  H_0 &=& -w_h\Big[\sum_{n=1}^{L_S-1}+\sum_{n=L_S+1}^{2L_S-1}\Big](c^{\dg}_{n+1}\tau_zc_{n}+{\rm h.c.}) \nn \\ && -\mu\sum_{n=1}^{2L_S}c^{\dg}_{n}\tau_zc_{n} + \De \sum_{n=1}^{L_S-1}c^{\dg}_{n+1}[\cos{(\phi_S)}\tau_x \nn \\ && +\sin{(\phi_S)}\tau_y]c_n + {\rm h.c.}  +\De\sum_{n=L_S+1}^{2L_S-1}(c^{\dg}_{n+1}\tau_xc_n \nn \\ &&+ ~{\rm h.c.}) -w_J(c^{\dg}_{L_S+1}\tau_zc_{L_S}+{\rm h.c.}), \nn \\ 
  H_{1}(t)&=&V_0\cos{\om t}\sum_{n=1}^{L_S}c^{\dg}_{n}\tau_zc_{n} \nn \\ && + V_0\cos{(\om t+\phi)} \sum_{n=L_S+1}^{2L_S}c^{\dg}_{n}\tau_zc_{n},  \label{eq:H}
 \eea
 where  $\phi_S$ is the superconducting phase difference between the two Kitaev chains, $w_J$ is the hopping amplitude at the bond that connects the two Kitaev chains to make the junction, and $\phi$ is the difference between the phases of the driving potentials on the two Kitaev chains. The terms with summation over $n=1,2,..,L_S-1$ describe the superconductor on the left, and the terms with summation over $n=L_S+1,..,2L_S-1$ describe the superconductor on the right. The superconducting phase difference $\phi_S$ is responsible for the Josephson effect in the undriven junction. On the other hand, $\phi$ is the difference between the phases of the driving potentials on the two Kitaev chains.  
 
 Floquet states $|v_j\ra$ are the eigenstates of the unitary time evolution operator $U_T$ that takes a  state of the system at time $t=0$ to the state at time $t=T$. While constructing $U_T$, the  time interval $[0,T]$ is sliced into $M$ equal intervals and the  Hamiltonian $H(t)$ is taken to be constant in each of these intervals, as described in earlier works~\cite{soori20pump,soori22sde}. 
The system is in the ground state of $H_0$ at time $t=-\infty$. The periodic driving is switched on over a timescale $\tau$ in such a way that the Hamiltonian of the system for $t<0$ is $H(t)=H_0+\eta(t) H_1(t)$, where $\eta(t)=e^{-t^2/\tau^2}$. The limit $\tau\to0$ refers to the periodic driving being switched on suddenly at $t=0$. At $t=-\infty$, the single particle eigenstates of $H_0$: $|u_i\ra$ for $i=1,2,..,N/2$ ($N=4L_S$) are occupied, and they time evolve into states $|\psi_i\ra$ at time $t=0$.

 The current operator at the junction between two superconductors is 
  \beq \hat J = \f{-iew_J}{\hbar}(c^{\dg}_{L_S}c_{L_S+1}-c^{\dg}_{L_S+1}c_{L_S}), 
 \label{eq:curr-op}\eeq 
 The current averaged over infinite time starting from $t=0$ is then given by
\bea J_{av} &=& \sum_{i=1}^{N/2} \sum_{j=1}^N |c_{i,j}|^2 (J_T)_{jj},
{\rm ~~where~~} c_{i,j} =  \langle \psi_i|v_j\rangle \nn \\ 
&{\rm and} & ~~(J_T)_{jj} = \f{1}{T}\sum_{k=1}^M\langle v_j|U^{\dg}(t_k,
0)|\hat J|U(t_k,0)|v_j\rangle dt,~~~~ \label{eq:curr} \eea
where $dt=t_k-t_{k-1}$,  the interval $[0,T]$ is sliced into $M$ sub-intervals of equal size and $t_k$ is the centre of the $k$-th sub-interval.

  \section{Floquet end modes}
 \begin{figure}[htb]
  \includegraphics[width=8cm]{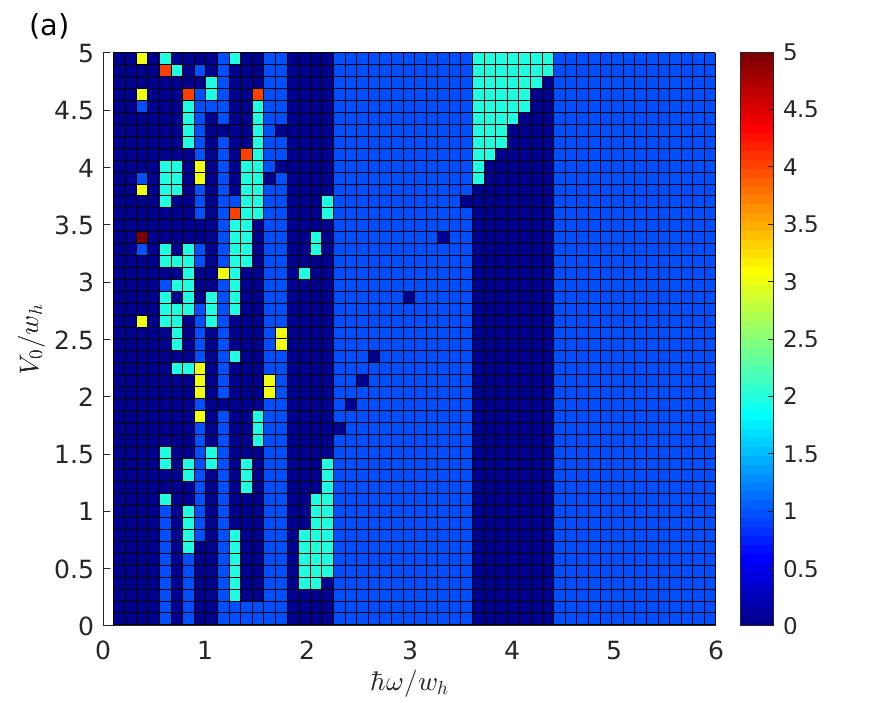}
  \includegraphics[width=8cm]{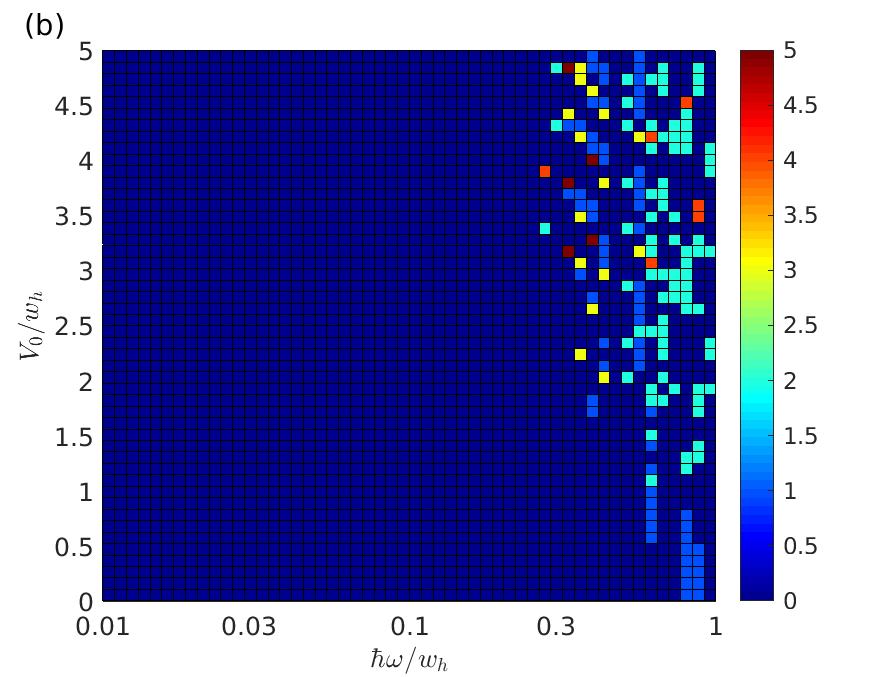}
  \caption{Winding number, which gives the  number of pairs of Majorana fermions in a long driven Kitaev chain. The range of $\hbar\om/w_h$ is different in the two plots. Also, in (a), $\om$ is plotted in linear scale whereas in (b), $\om$ is in log-scale. Parameters: $\mu=0.2w_h$, $\De=0.9w_h$ and  $M=80$.  }\label{fig:wind}
 \end{figure}
 
 The Kitaev chain with periodically varying chemical potential can host Floquet Majorana fermions. The number of Floquet Majorana fermions can be calculated by the winding number.  Due to bulk-boundary correspondence, a knowledge of the topology of the bulk bands in a system with periodic boundary condition is sufficient to discern the number of topologically protected edge states in an open system~\cite{benito14}. Winding number gives the  number of pairs of Majorana fermions in a long driven open  Kitaev chain~\cite{thakurathi13}. The unitary time evolution operator $U_{T,k}$ that takes the system from time $t=0$ to $t=T$ at momentum $\hbar k$ for a translationally invariant system can be expressed as $U_{T,k}=e^{-ih_{eff}(k)T}$, where $h_{eff}(k)$ is a $2\times 2$ matrix having  the form $h_z(k)\tau_z+h_x(k)\tau_x$. The number of times the loop in $(h_x,h_z)$ encircles the origin $(0,0)$ as $k$ is varied from $-\pi$ to $\pi$ is the winding number. The Floquet Majorana fermions in  an open system come in two varieties, having the Floquet eigenvalues $1$ and $-1$.  The winding number gives the sum of the total number of pairs of Majorana fermions belonging to both these varieties. In Fig.~\ref{fig:wind}, the winding number is numerically calculated and plotted in a color plot as a function of the driving frequency $\om$ and amplitude of driving potential $V_0$ for $\mu=0.2w_h$, $\De=0.9w_h$, $M=80$.

\begin{figure}
  \includegraphics[width=8cm]{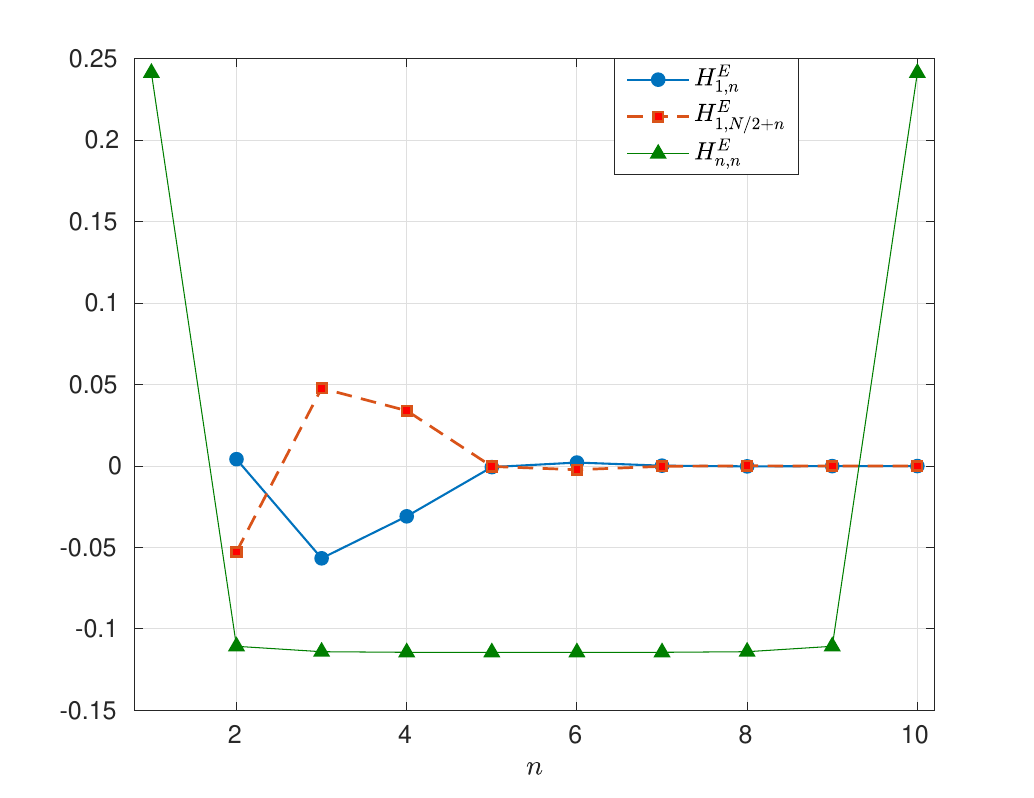}
  \caption{Terms in the effective Hamiltonian of the Floquet system in units of $w_h$ for $\hbar\om=2w_h$, $V_0=0.5w_h$, $L_S=10$,  $\mu=0.2w_h$, $\De=0.9w_h$ and  $M=80$. $H^E_{1,n}$ is the electron hopping amplitude between sites $1$ and $n$. $H^E_{1,N/2+n}$ is the superconducting hopping between sites $1$ and $n$. $H^E_{n,n}$ is the onsite potential.}\label{fig:Heff}
 \end{figure}

 For certain choice of parameters, the Floquet end states turn out to be non-topological anomalous end modes. These modes have Floquet eigenvalues much different from $1$ and $-1$. Such modes were predicted in Kitaev chains where the nearest neighbor hopping amplitudes are periodically driven~\cite{saha17}. For $\hbar\om=2w_h$, $V_0=0.5w_h$, $\mu=0.2w_h$, $\De=0.9w_h$, $M=80$ and $L_S=10$,  we find that in addition to two pairs of Floquet Majorana modes, two pairs of anomalous end modes exist with Floquet eigenvalues close to $e^{\pm i 0.3\pi}$. The matrix elements of the  effective Hamiltonian $H^E$ which satisfies  $U_T=e^{-iH^ET}$ can give an insight into the mechanism behind the appearance of anomalous Floquet end modes.  We find the effective Hamiltonian numerically. From the effective Hamiltonian, we plot the hopping amplitudes between sites $1$ and $n$: $H^E_{1,n}$, the superconducting hopping amplitudes between sites $1$ and $n$: $H^E_{1,N/2+n}$ and the onsite energies $H^E_{n,n}$ versus $n$ in Fig.~\ref{fig:Heff}. 
It can be seen that in the effective Hamiltonian, there is a large onsite energy at the ends of the chain. Further, the electron hopping and superconducting hopping amplitudes have substantial magnitude for next-to-next and next-to-next-to-next neighbor sites. This explains the existence of non-topological anomalous end modes localized at the ends of the chain. 
 
 \section{Adiabatic preparation of Floquet states}
In this section, we  study the states of a single Kitaev chain driven periodically. Floquet states are characterized by quasienergy which is periodic with a period  $\hbar\om$ in contrast to the equilibrium systems which have eigenenergy. This makes it difficult to define a ground state and the occupation of the Floquet states. To overcome this problem, the periodic in time terms in the  Hamiltonian are switched on gradually starting from the ground state of the  equilibrium system~\cite{kitagawa11,heinisch16,bandho19,bandho21}. An eigenstate of the equilibrium system  evolves into a linear combination of different Floquet states, and the overlap of the initial state with different Floquet states determines the long time averaged expectation value of a physical quantity of interest. To quantify the extent to which an initial state $|u_{K,i}\ra$ has overlap with different Floquet states $|v_{K,j}\ra$, we define inverse participation ratio (IPR): $I_i=\sum_{j}|\la v_{K,j}|u_{K,i}\ra|^4$. IPR takes values between $0$ to $1$ and a value equal to $1$ means that the initial state has time evolved into only one of the Floquet states, whereas a much smaller value of IPR implies that the initial state evolves into  a linear combination of many Floquet states. We average over all the initially occupied states $|u_{K,i}\ra$ to quantify the extent to which the system has evolved adiabatically into Floquet states: $I_{av}=\sum_{i=1}^{N/2}2I_i/N$. 
\begin{figure}[htb]
 \includegraphics[width=4.0cm]{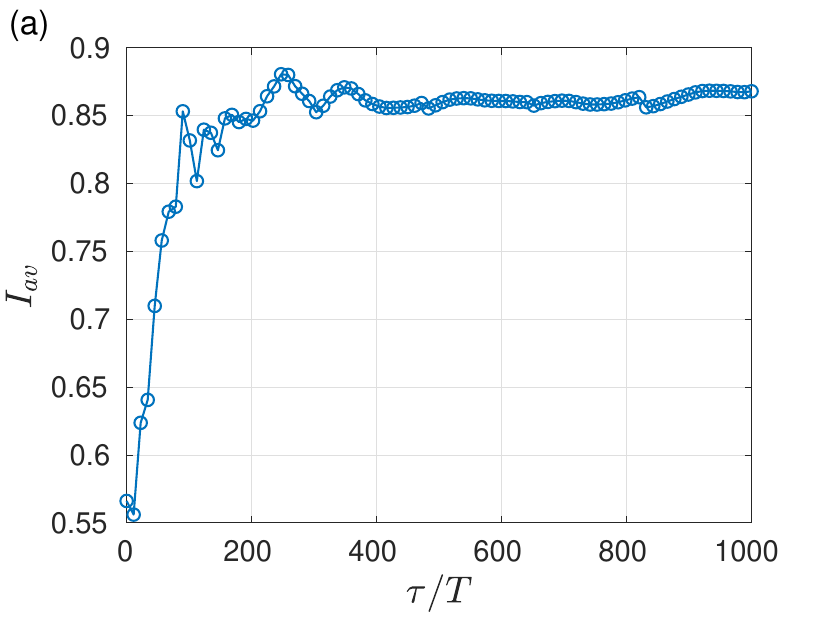}
 \includegraphics[width=4.0cm]{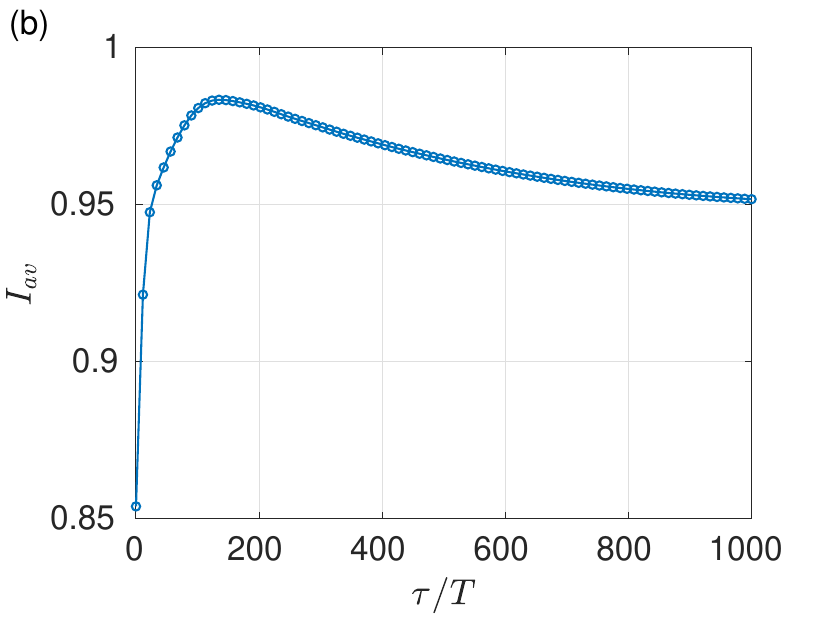}
 \includegraphics[width=4.0cm]{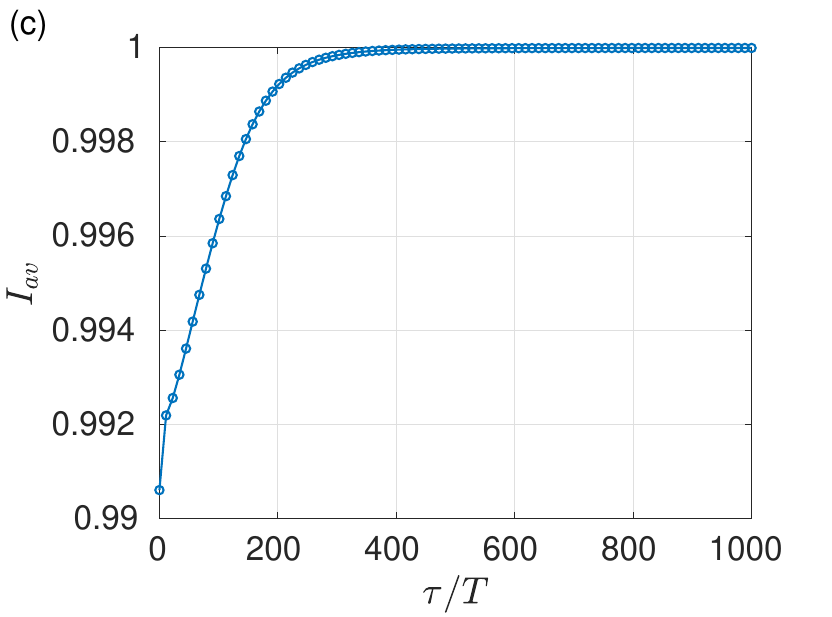}
 \includegraphics[width=4.0cm]{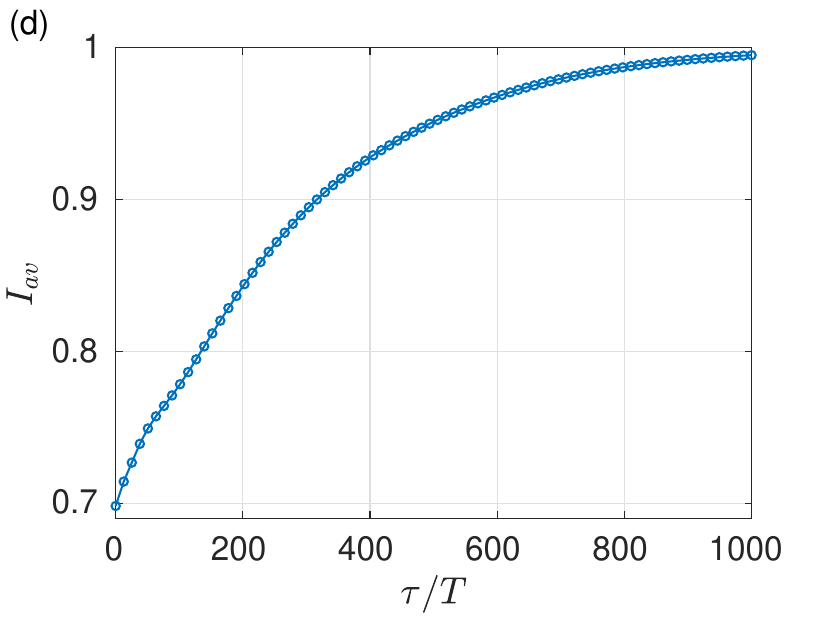}
 \caption{Average inverse participation ratio $I_{av}$ versus the timescale of slow evolution for (a)$\hbar\om=0.01w_h$, (b)$\hbar\om=0.1w_h$, (c)$\hbar\om=1w_h$ and (d)$\hbar\om=1.8w_h$. Other parameters: $L_S=10$, $\mu=0.2w_h$, $\De=0.9w_h$, $V_0=0.5w_h$ and  $M=80$.  A value of $I_{av}=1$ implies that the initial state has adiabatically evolved into one of the Floquet states. } \label{fig:ipr}
\end{figure}
We start with the ground state of the Hamiltonian $H_{K,0}$ at time $t=-\infty$ and switch on the periodic driving so that the full Hamiltonian of the system is $H=H_{K,0}+\eta(t)H_{K,1}(t)$, where $\eta(t)=e^{-t^2/\tau^2},$ for $t<0$ and $\eta(t)=1,$ for $t\ge 0$. $H_{K,1}(t)$ is periodic in time with a frequency $\om$.  We find that $I_{av}$ is large for larger values of $\tau/T$. Also, we find that for small values of $\om$, $I_{av}$ is lower, since the driving can take the states away from the occupied states to nearby eigenstates of $H_{K,0}$ more easily. However, as $\om$ increases, for a state with energy $E$ near the gap,  there are no plane wave states of $H_{K,0}$ at energies $E\pm\hbar\om$ and the mixing between the states is suppressed  making  $I_{av}$ larger. Further, for values of  $\hbar\om$ close to $E_g=\De\sqrt{(4w_h^2-\mu^2-4\De^2)/(w_h^2-\De^2)}$ (where $2E_g$ is the superconducting gap), mixing between the levels mediated by the Majorana bound state results in a lower value of $I_{av}$. These features can be seen in Fig.~\ref{fig:ipr}. For the choice of parameters in Fig.~\ref{fig:ipr}, $E_g=1.752w_h$. Hence, the IPR is lower at $\hbar\om=1.8w_h$. 

\section{Current following swift switch}

\begin{figure}[htb]
 \includegraphics[width=2.8cm]{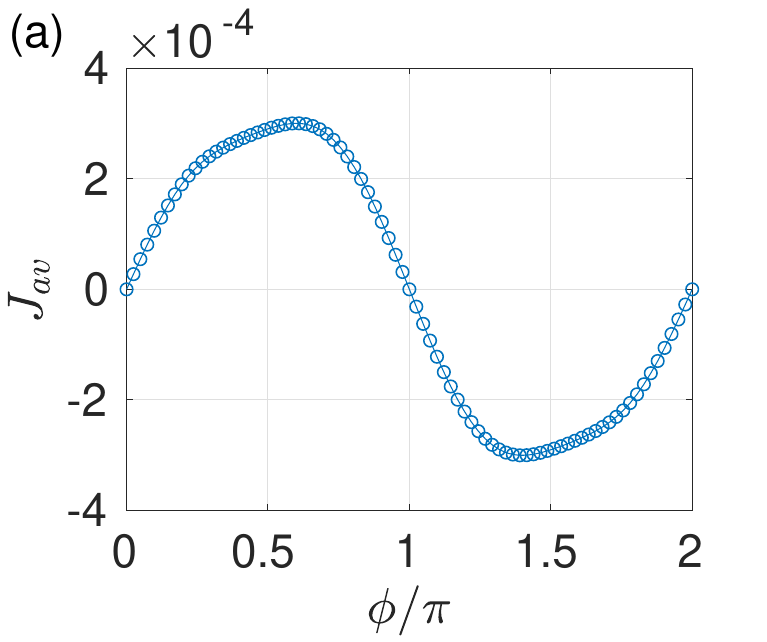}
 \includegraphics[width=2.8cm]{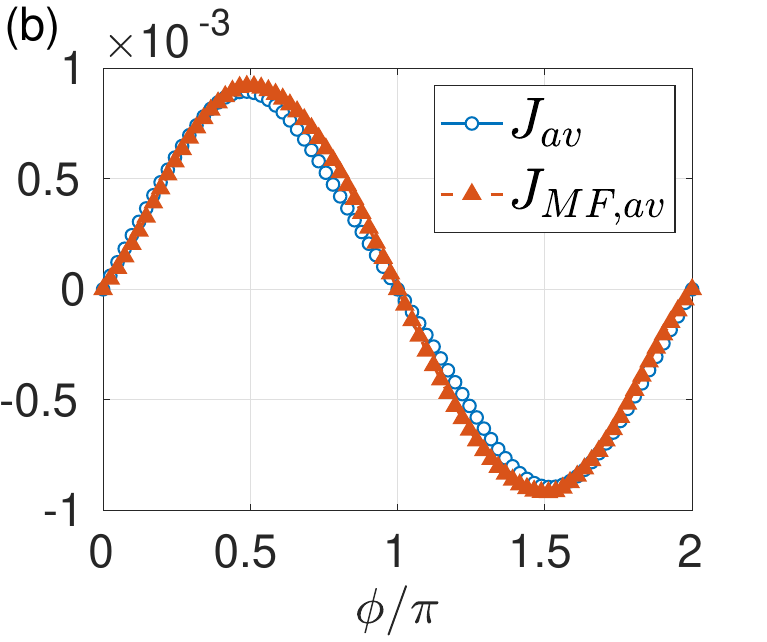}
 \includegraphics[width=2.8cm]{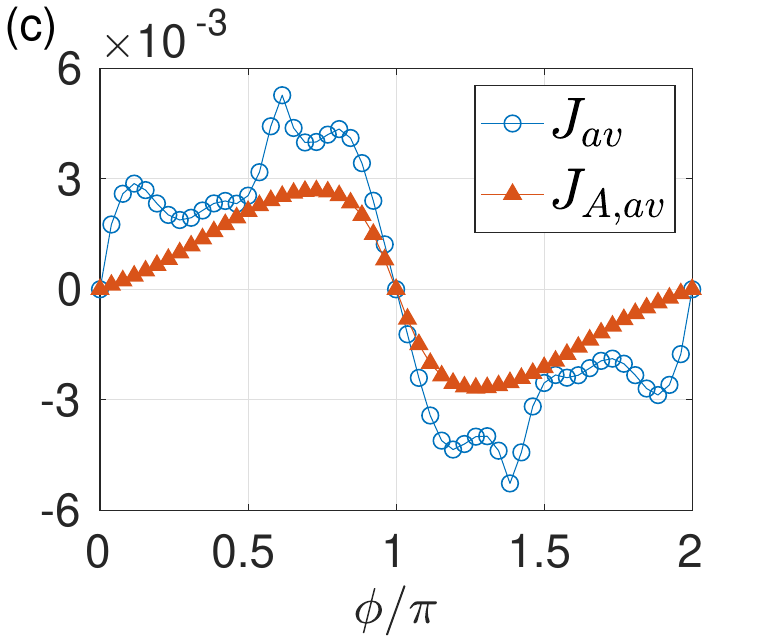}
 \caption{Long time averaged current in units of $ew_h/\hbar$ versus the difference in phases of the driving potential in absence of a superconducting phase difference. (a) $\hbar\om=0.5w_h$,   (b) $\hbar\om=3w_h$, (c) $\hbar\om=2w_h$. Other parameters: $\mu=0.2w_h$, $w_J=0.2w_h$, $\De=0.9w_h$, $V_0=0.5w_h$,  $M=80$ and $L_S=10$.  For (a), the winding number is $0$ and for (b), the winding number is $1$. In (b), the weighted current $J_{MF,av}$ that signifies the contribution from Floquet Majorana fermion is also plotted. In (c), the weighted current $J_{A,av}$ signifies the current carried by the anomalous end modes.}\label{fig:psje}
\end{figure}
In this section, we  study the current at the junction between two Floquet superconductors after the periodic driving is switched on swiftly at time $t=0$. For $t<0$, the system is in ground state of the undriven Hamiltonian. At time $t=0$, periodic driving is switched on suddenly, and we calculate the long time averaged current at the junction between two driven Kitaev chains. To begin with, we choose the superconducting phase difference between the Kitaev chains to be zero, and we drive the two Kitaev chains with  time dependent chemical potentials that have the forms: $\mu-V_0\cos \om t$ and  $\mu-V_0\cos (\om t+\phi)$. In Fig.~\ref{fig:psje}, the long time averaged current versus the difference in the phase of the driving potential $\phi$ is plotted for three  values of $\om$. It is evident from this figure that a net average  current can be driven across a junction between two Floquet superconductors if the difference between  phases of  the driving potential is nonzero even when the superconducting phase bias is absent. This is one of the important results of this work, and we term such a current anomalous current. For the choice  $\hbar\om=0.5w_h$, $V_0=0.5w_h$, $\mu=0.2w_h$ the winding number of the constituent Floquet superconductors is $0$. For the choice  $\hbar\om=3w_h$, $V_0=0.5w_h$, $\mu=0.2w_h$ the winding number of the constituent Floquet superconductors is $1$. 

To get an idea of the contribution of the Floquet Majorana fermions to the time average current, we calculate a weighted long time averaged current, giving higher weight to the localized Floquet states in the following way. For a given value of $\phi$, let $I_j$ be the IPR of the $j$-th Floquet state,  let $I_0$ be the maximum value of IPR among the Floquet states of the system, and let $J_{j}$ be the contribution of $j$-th Floquet state to the long time averaged current. Then, $J_{MF,av}=\sum_je^{-100(I_0 - I_j)}J_j$ quantifies the current carried by Floquet states according to the extent to which they are localized. The contribution from the  Floquet states with value of IPR lower than $I_0$ are exponentially suppressed. In Fig.~\ref{fig:psje}(b), the weighted long time averaged current $J_{MF,av}$ is plotted as a function of $\phi$ for the choice of parameters $\hbar\om=3w_h$  wherein each  Floquet superconductor has one pair of Majorana fermions. The Floquet states with large IPR here are Floquet Majorana end states, as indicated by their Floquet eigenvalues.  It can be inferred from Fig.~\ref{fig:psje}(b) that the Floquet Majorana fermions contribute significantly to the long time averaged current. In Fig.~\ref{fig:psje}(c), the weighted long time averaged current $J_{A,av}$ which has the same expression as that of $J_{MF,av}$ is plotted along with $J_{av}$ for $\hbar\om=2w_h$. For this choice of parameters, there are anomalous Floquet end modes in the system and the current $J_{A,av}$ is the current carried by the anomalous end modes as indicated by the Floquet eigenvalues for these states. This shows that even the anomalous end modes contribute significantly to the current whenever present. 

\begin{figure}[htb]
 \includegraphics[width=7cm]{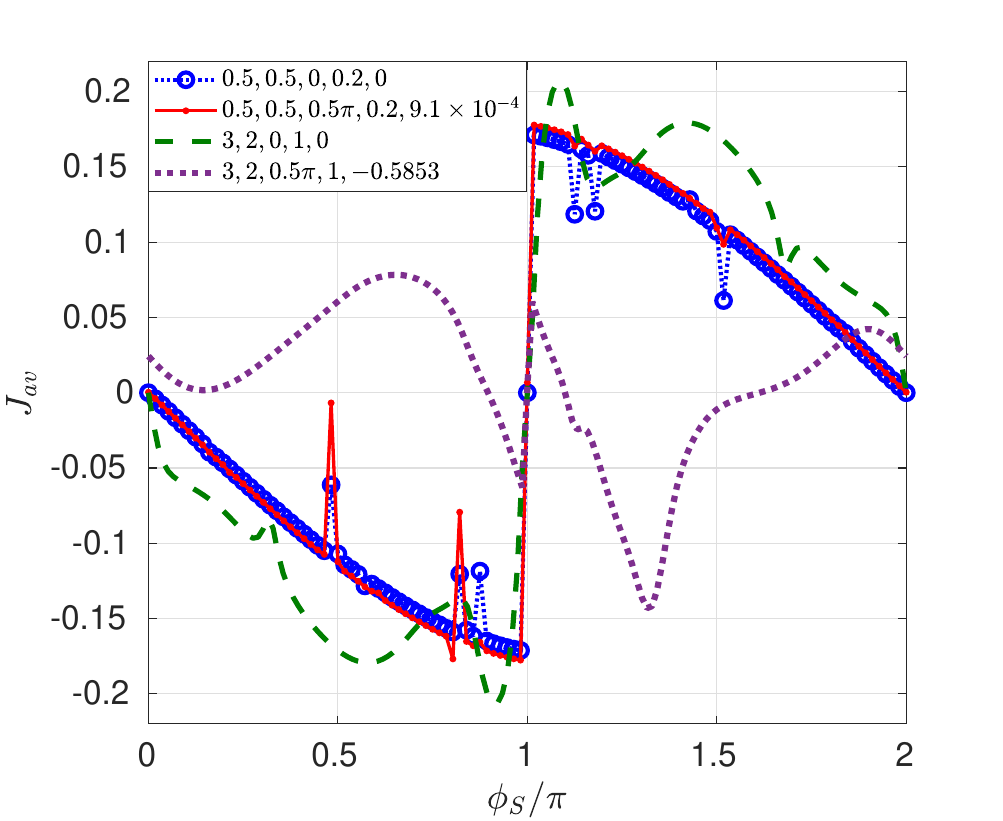}
 \caption{Current phase relation of the Josephson junction between Floquet superconductors. Current is in units of $ew_h/\hbar$. Parameters: $\mu=0.2w_h$, $M=80$ and $L_S=10$. In the legend, values of $\hbar\om/w_h, V_0/w_h,\phi, w_J/w_h, \ga$ are indicated for each curve. }\label{fig:cpr}
\end{figure}

\begin{figure}[htb]
 \includegraphics[width=4.2cm]{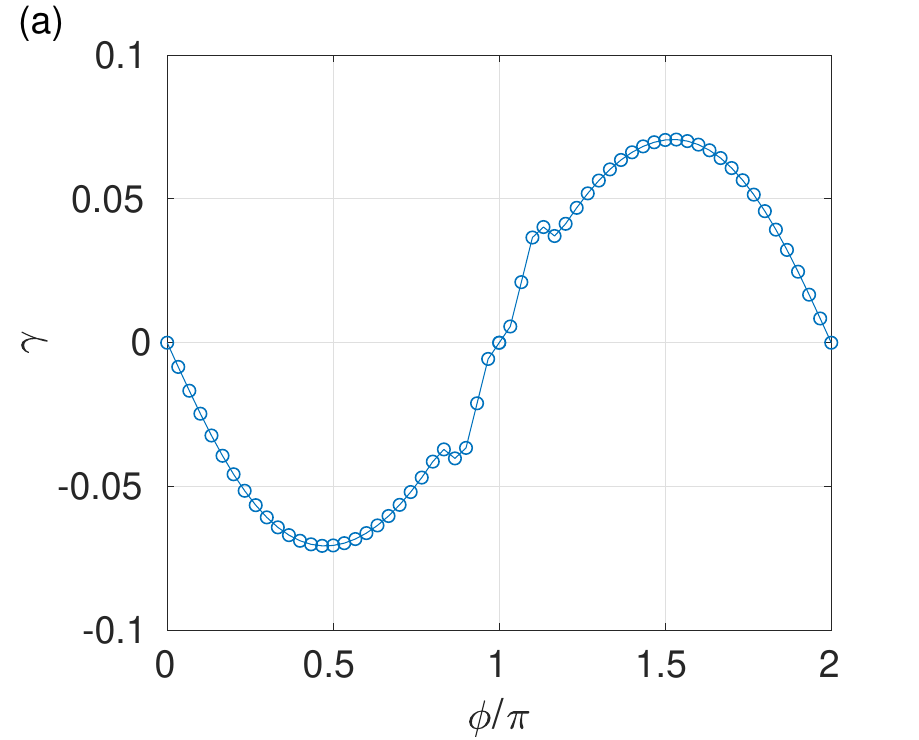}
 \includegraphics[width=4.2cm]{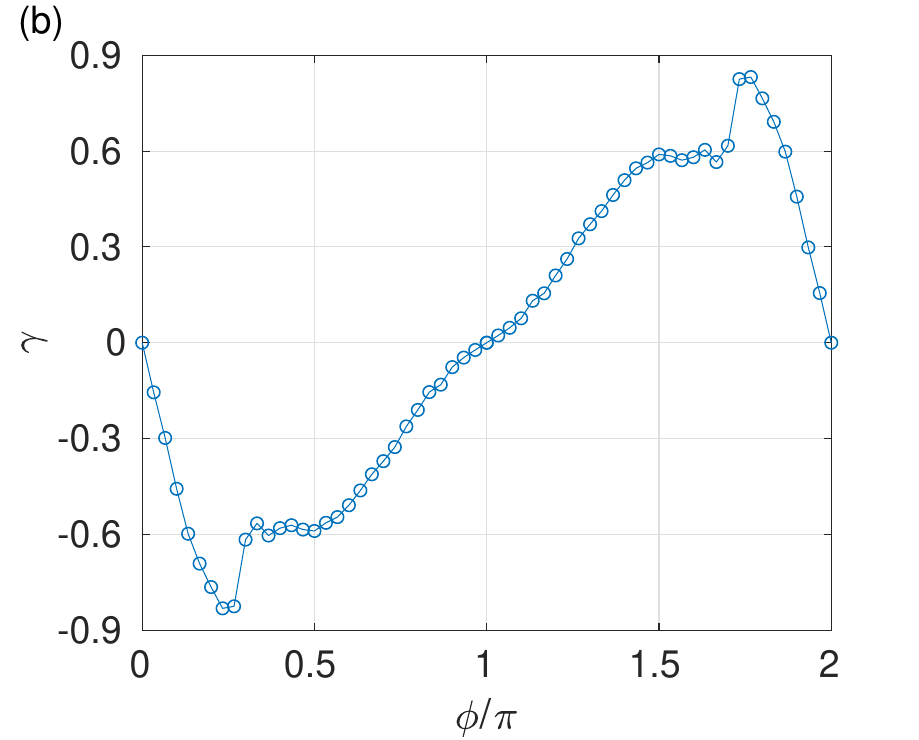}
 \caption{Diode effect coefficient versus the difference in phases of the driving potential for (a) $V_0=0.5w_h$, $w_J=0.2w_h$, (b) $V_0=2w_h$, $w_J=w_h$. Other parameters: $\hbar\om=3w_h$,  $\mu=0.2$, $\De=0.9w_h$, $L_S=10$ and  $M=80$. }\label{fig:ga}
\end{figure}

Now, we turn to the behavior of long time averaged current when a superconducting phase difference $\phi_S$ is applied between the Floquet superconductors. We investigate the current phase relation at different values of $\phi$. In Fig.~\ref{fig:cpr}, the current phase relation is plotted for different choices of $\om, V_0, \phi, w_J$. In a current phase relation, the maximum and minimum values of the current not being equal in magnitude  signals a diode effect. The diode effect is quantified by the  diode effect coefficient, defined by $\ga=2(I_{max}+I_{min})/(I_{max}-I_{min})$ where $I_{max}$ and $I_{min}$ are the maximum and minimum values of the current respectively in the current phase relation. The legend indicates the values of $\hbar\om/w_h, V_0/w_h, \phi, w_J/w_h, \ga$ for each curve. This shows that for smaller values of $\om$ and $V_0$, even though the Josephson current is large in magnitude,  a nonzero value of  $\phi$ does not lead to substantial diode effect with a value of $\ga=9.1\times 10^{-4}$ for $\phi=\pi/2$. For $\hbar\om=3w_h$, $V_0=0.5w_h$, $w_J=w_h$, and $\phi=\pi/2$, the diode effect coefficient is substantially larger in magnitude with a value $-0.07$.    For the choice $(\hbar\om, V_0, w_J)=(3w_h,0.5w_h,w_h)$, Floquet Majorana fermions participate in the transport and contribute significantly to the current. This means the Floquet Majorana fermions significantly contribute to the  diode effect. We set $\hbar\om=3w$ and plot the diode effect coefficient $\ga$ versus $\phi$ in Fig.~\ref{fig:ga}. In Fig.~\ref{fig:ga}(a), $V_0=0.5w_h$ and $w_J=0.2w_h$, and the diode effect coefficient reaches a maximum value of around 0.07 while in Fig.~\ref{fig:ga}(b), $V_0=2w_h$ and $w_J=w_h$, and the diode effect coefficient can be as large as $0.85$. This shows that the diode effect coefficient can be made large by choosing a higher value of driving amplitude $V_0$ and making the junction transparent (by choosing $w_J=w_h$). 

\section{Current following slow switch}
\begin{figure}[htb]
 \includegraphics[width=4.2cm]{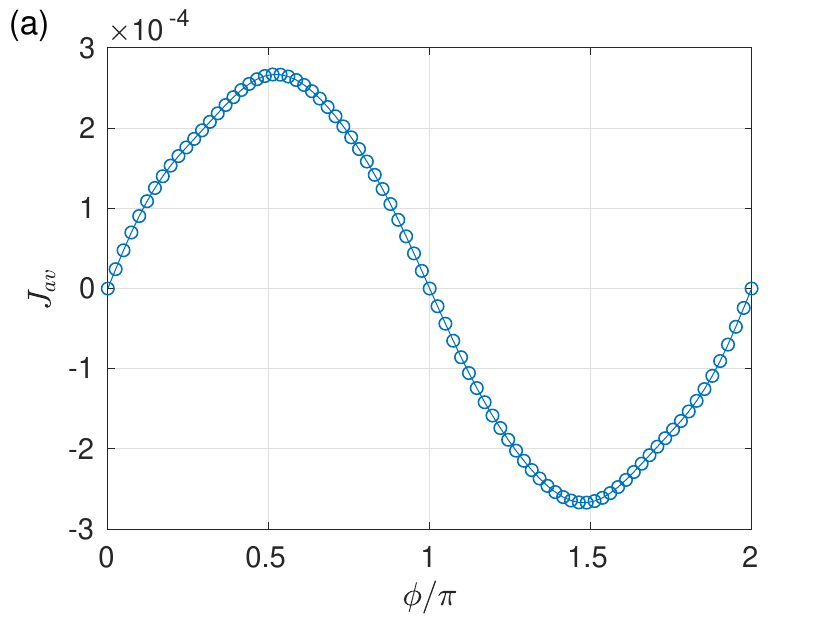}
 \includegraphics[width=4.2cm]{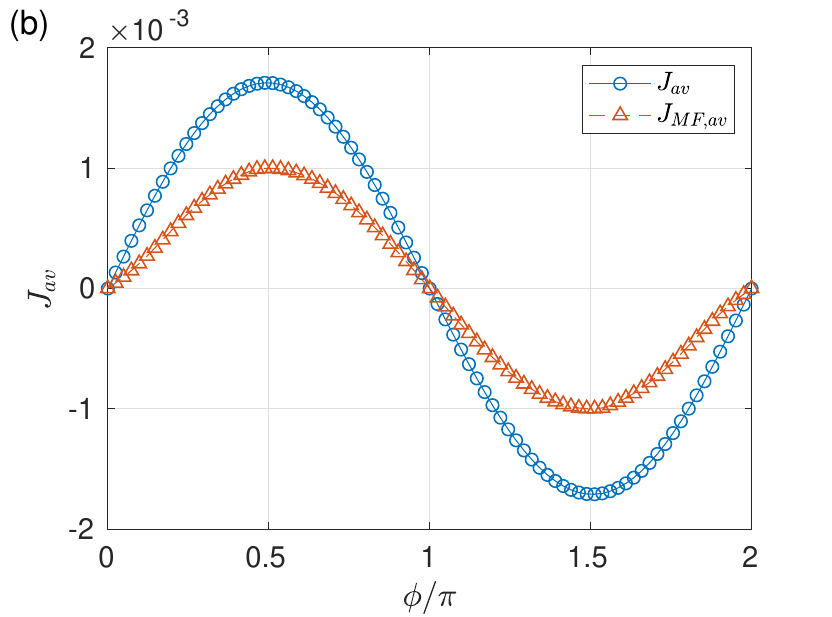}
 \caption{Time averaged current between two Floquet superconductors maintained at the same superconducting phase in units of $ew_h/\hbar$ versus the difference in phase of the driving potential $\phi$ after switching on the periodic potential adiabatically. The timescale $\tau$ over which the periodic driving is switched on is $100T$. (a) $\hbar\om=0.5w_h$ and  (b) $\hbar\om=3w_h$.  Other parameters: $\mu=0.2w_h$, $w_J=0.2w_h$, $\De=0.9w_h$, $V_0=0.5w_h$,  $M=80$ and $L_S=10$. In (b), the curve with triangle shaped data points is the weighed current that gives higher weight to the current carried by the Floquet Majorana modes. }\label{fig:psje-adia}
\end{figure}

\begin{figure}
 \includegraphics[width=4.2cm]{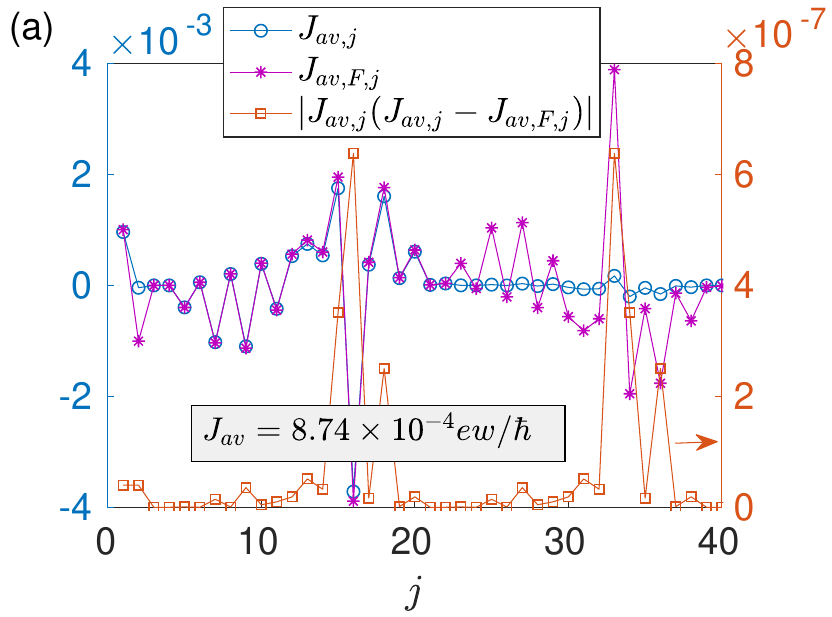}
 \includegraphics[width=4.2cm]{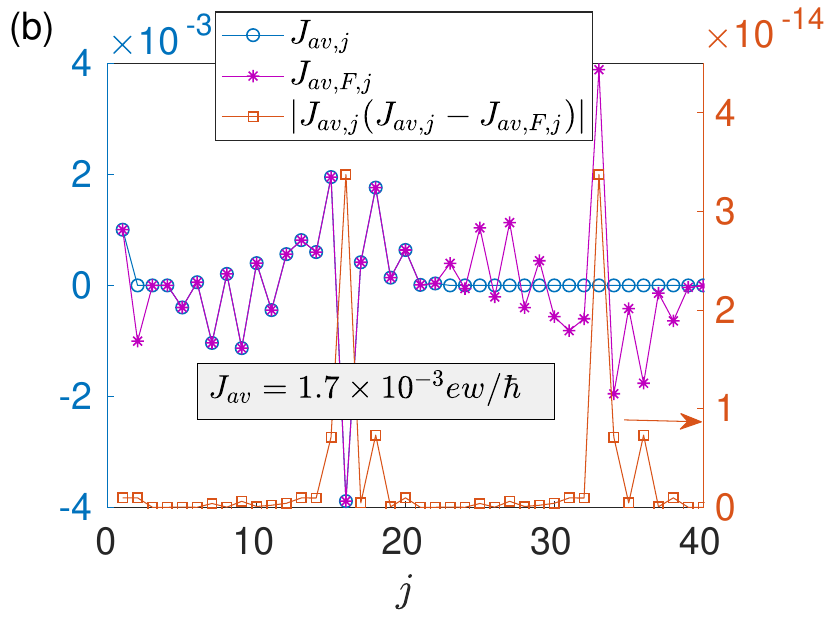}
 \caption{Current contribution Floquet states to the long time averaged current and the current carried by the individual Floquet states (on the left ordinate). The unit of current is $ew_h/\hbar$.  The product $|J_{av,j}(J_{av,j}-J_{av,F,j})|$ is  on the right ordinate. (a) $\tau=0.1T$, $J_{av}=-8.74\times 10^{-4} ew_h/\hbar$ (b) $\tau=100 T$, $J_{av}=1.7\times10^{-3}ew_h/\hbar$.  Other parameters: $\hbar\om=3w_h$, $\mu=0.2w_h$, $w_J=0.2w_h$, $\De=0.9w_h$, $V_0=0.5w_h$,  $M=80$ and $L_S=10$. }\label{fig:Flo-curr}
\end{figure}

In this section, we shall study the behavior of the long time averaged current $J_{av}$ at the junction when the periodic drive is switched on adiabatically. To begin with, we look at the dependence of $J_{av}$ on $\phi$- the difference in phases of the driving potentials of the two Floquet superconductors, having no difference between their superconducting phases. Similar to Fig.~\ref{fig:psje}, we plot $J_{av}$ versus $\phi$ for the two cases: (a) $\hbar\om=0.5w_h$, and  (b) $\hbar\om=3w_h$ in Fig.~\ref{fig:psje-adia} choosing other parameters:  $\mu=0.2w_h$, $w_J=0.2w_h$, $\De=0.9w_h$, $V_0=0.5w_h$,  $M=80$ and $L_S=10$. In this figure, the periodic in time driving is switched on adiabatically with a Gaussian envelope over a timescale $\tau=100T$. On comparing Fig.~\ref{fig:psje-adia} with Fig.~\ref{fig:psje}, it can be seen that for $\hbar\om=0.5w_h$, switching on the periodic driving adiabatically does not make much difference to the value of $J_{av}$. For $\hbar\om=3w_h$, the value of $J_{av}$ increases significantly on adiabatically switching on the periodic driving. This is because, the gap in the spectrum of the undriven system is approximately $3.5w_h$ and for a frequency $\hbar\om=0.5w_h$, the extent of mixing between the energy levels is lower compared to that in the case $\hbar\om=3w_h$. For $\hbar\om=0.5w_h$, the two Floquet superconductors do not host Floquet Majorana modes. For $\hbar\om=3w_h$, each Floquet superconductor hosts a pair of Floquet Majorana modes. When the system is driven adiabatically, not only the Majorana modes, but the bulk modes also carry a significant fraction of the long time averaged current as can be seen from Fig.~\ref{fig:psje-adia}(b). 
To get a further insight into the magnitude of the long time average current in absence of a superconducting phase difference, we plot the currents carried by the individual Floquet states and their contribution to the long time averaged current in Fig.~\ref{fig:Flo-curr} for the two cases when the periodic driving is switched on suddenly and slowly. For each of the cases, the contribution $J_{av,j}$ of $j$-th Floquet state to the long time averaged  current and the current carried $J_{av,F,j}$ by the $j$-th Floquet state are plotted with ordinate on the left axis. On the right axis, the product $|J_{av,j}(J_{av,j}-J_{av,F,j})|$ is also plotted. The product $|J_{av,j}(J_{av,j}-J_{av,F,j})|$ is zero either when the Floquet state contributes  0\% or 100\% to the long time averaged current. It can be seen that when the periodic driving is adiabatically switched on, the product is minuscule ($\sim 10^{-14}$) for all the Floquet states whereas for sudden switching of the driving, the product is around $\sim 10^{-7}$ for the Floquet states that carry large current. The current carried by a pair of Floquet states is equal in magnitude and opposite in sign. Hence, a nonadiabatic switching of the driving leads to a longtime averaged current drawing non-negligible contributions from both these Floquet states resulting in a lower value of $J_{av}$. 

\begin{figure}[htb]
 \includegraphics[width=6cm]{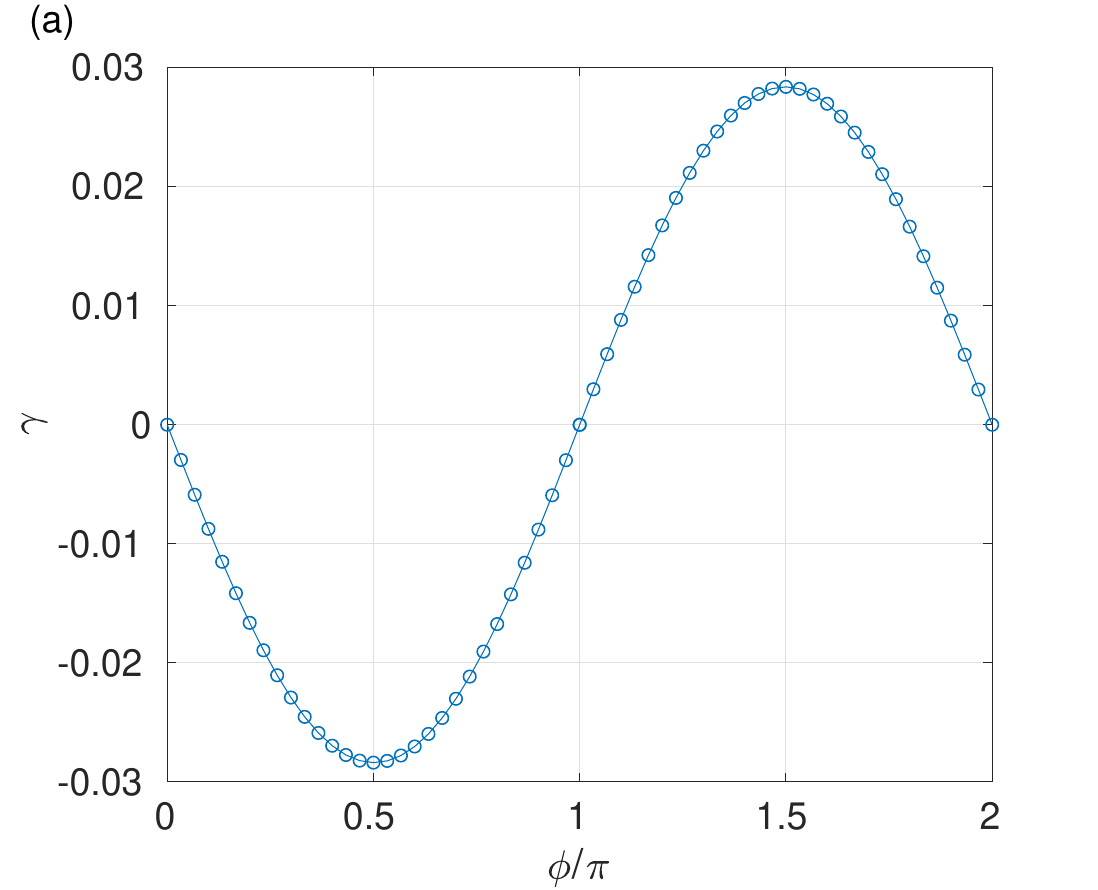}
 \includegraphics[width=4.2cm]{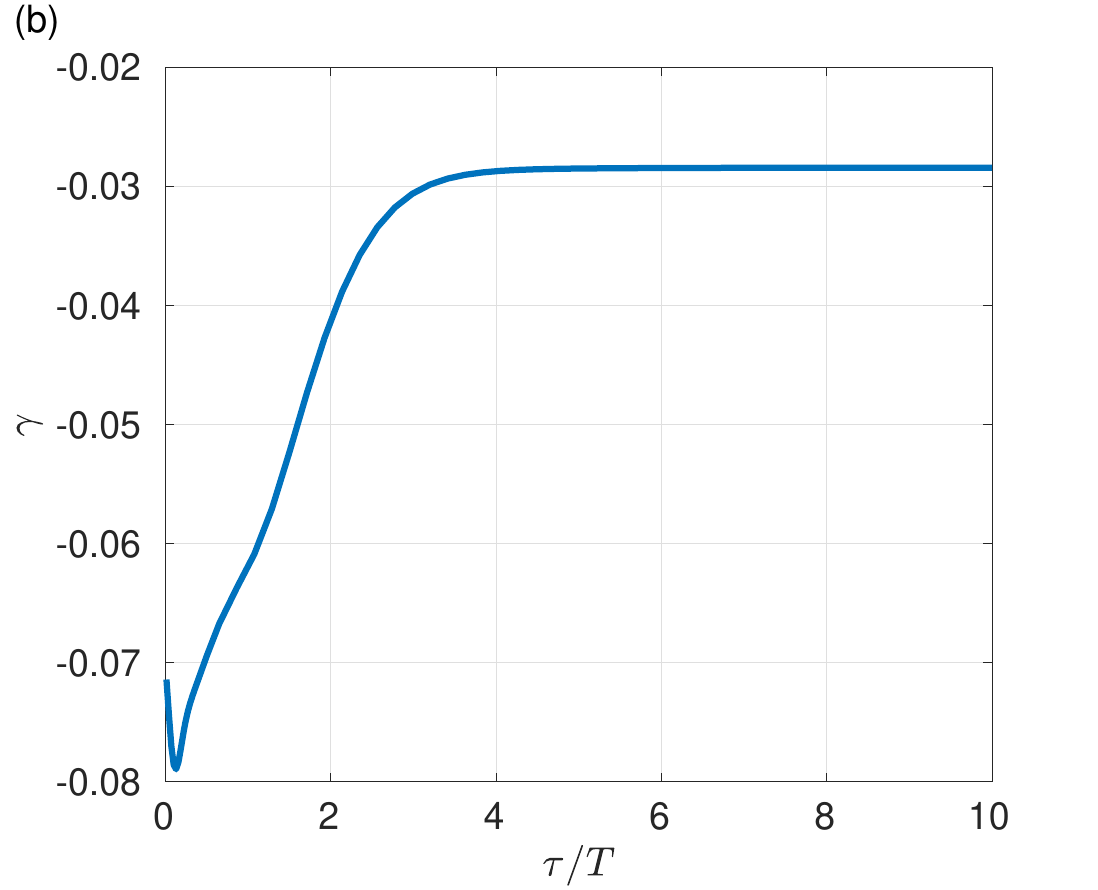}
 \includegraphics[width=4.2cm]{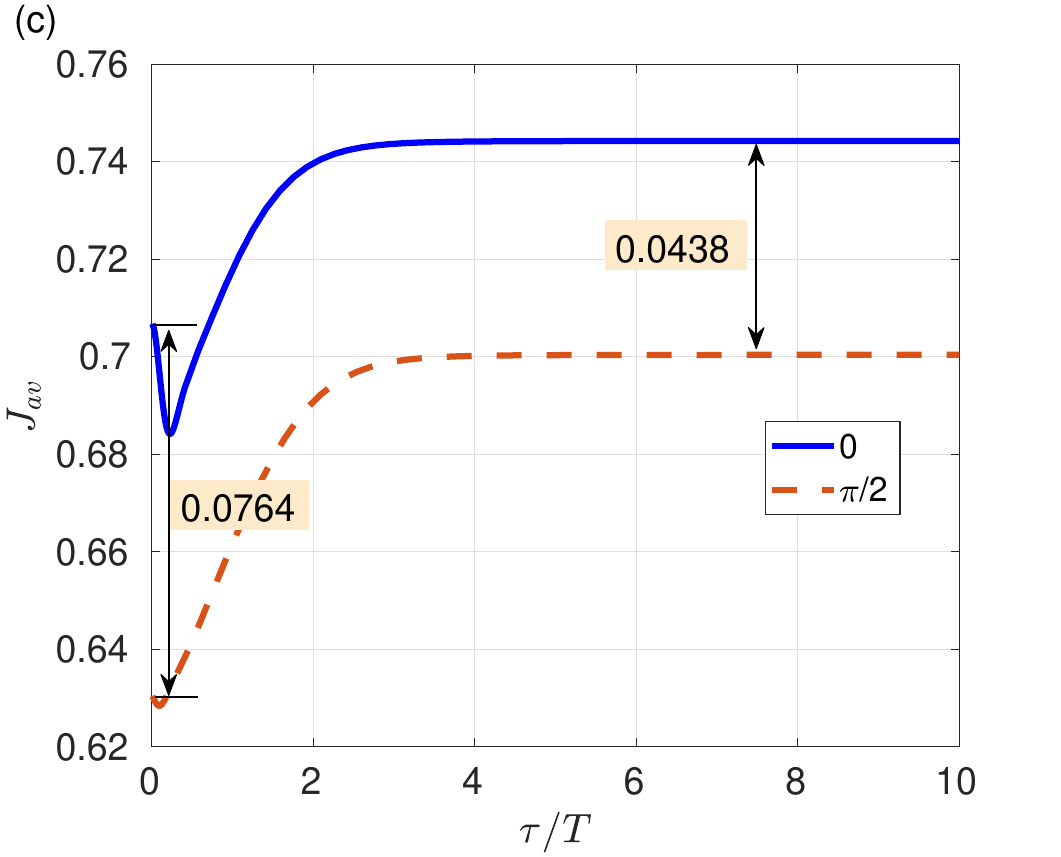}
 \caption{(a) Diode effect coefficient versus the difference in phases of the driving potential $\phi$ upon turning on the periodic driving slowly over a timescale $\tau=50T$, (b) Diode effect coefficient versus the timescale of switching on the periodic driving $\tau$, (c) Long time averaged Josephson current in units of $ew_h/\hbar$ versus timescale $\tau$ over which periodic driving is switched on for $\phi_S=1.001\pi$ plotted for two different values of $\phi$ indicated in the legend. Parameters: $\hbar\om=3w_h$, $\mu=0.2w_h$, $w_J=w_h$, $\De=0.9w_h$, $V_0=0.5w_h$,  $M=80$ and $L_S=10$. It is evident from (c) that the difference in the Josephson currents for $\phi=0$ and $\phi=\pi/2$ is much smaller for larger values of $\tau$. }\label{fig:ga-adia}
\end{figure}

We now turn to the current phase relation and the diode effect when the periodic driving in the system is slowly switched on. The current phase relation is qualitatively very similar to that in Fig.~\ref{fig:cpr}, except for a small change in the numerical values of the current. We find that the diode effect coefficient $\gamma$ decreases in magnitude upon switching on the driving adiabatically. In Fig.~\ref{fig:ga-adia}(a), we plot the diode effect coefficient versus $\phi$ for the choice of the timescale of switching on $\tau=50T$. In Fig.~\ref{fig:ga-adia}(b), we plot the diode effect coefficient versus the timescale of switching on $\tau$ for $\phi=\pi/2$. In Fig.~\ref{fig:ga-adia}(c), the Josephson current for the choice of superconducting phase difference $\phi_S=1.001\pi$ is plotted versus the timescale of switching on $\tau$ for $\phi=0, \pi/2$. It can be seen that the difference between the Josephson currents for $\phi=0$ and $\phi=\pi/2$ is much larger for sudden switching compared to that for adiabatic switching.
This is in contrast to the behavior of the long time averaged current driven by a nonzero $\phi$ in the absence of a superconducting phase difference. It has to be noted that the current between the Floquet superconductors due to a nonzero $\phi$ in absence of a superconducting phase difference is much smaller in magnitude ($\lesssim 10^{-3} ew_h/\hbar$) compared to the current due to  a superconducting difference ($\sim 0.74 ew_h/\hbar$ for $\phi_S=1.001\pi$). Further, the deviation of the Josephson current due to a nonzero value of  $\phi$ is much smaller when periodic driving is switched on adiabatically in contrast to that in absence of a superconducting phase difference. It can be seen from Fig.~\ref{fig:ga-adia}(c) that the difference between Josephson currents for $\phi=0$ and $\phi=\pi/2$ is $0.0764ew_h/\hbar$ for sudden switching, while it is $0.0438ew_h/\hbar$ for $\tau>3T$. This explains why the diode effect coefficient has a lower value when the driving is  adiabatically switched on. 

\section{Summary and Conclusion}
We studied the Floquet states of a periodically driven Kitaev chain and calculated the winding number of translationally invariant chain as a function of the driving frequency and amplitude of the driving potential. We also find that a driven  open Kitaev chain hosts  end modes that can be either topological Floquet Majorana modes or anomalous non-topological end modes, depending on the choice of parameters. We explain the reason behind anomalous Floquet end modes, and their origin cannot be explained using the topology characterized by the winding number. In Floquet systems, it is not possible to define a ground state. We overcome this problem by  starting with the ground state of the undriven system and switching on the periodic driving gradually over a timescale $\tau$. The  dynamics of the system is dictated by the overlap of different  Floquet states of the system with states of the system occupied once the periodic driving is completely switched on. We calculate inverse participation ratio and characterize the extent to which the single particle states of the equilibrium system get mapped to Floquet states. We find that switching on the periodic driving gradually over a long timescale results in initially occupied single particle states of the undriven system evolving into a linear combination of Floquet states with significant weights on fewer Floquet states.  In a junction between two periodically driven superconductors, a long time averaged current can flow in the absence of a phase bias if the driving potentials of the two superconductors differ by a phase. We find that the Floquet Majorana end modes and anomalous Floquet end modes contribute significantly to the long time averaged current whenever present.  Further, we study the current phase relation and Josephson diode effect when the two Floquet superconductors that form the junction  are driven with a difference between the phases of the driving potential.  The anomalous current due to a difference in phases of the driving potential in absence of a superconducting phase difference  is much smaller in magnitude compared to the long time averaged  Josephson current between the Floquet superconductors.  We find that the current in absence of superconducting phase difference is substantially large for the  difference $\phi=\pi/2$ between phases of the driving potentials, and the magnitude of this current increases with the timescale of switching on the periodic driving. In contrast, we find that the diode effect coefficient, which is substantial for $\phi=\pi/2$  decreases with the timescale $\tau$ over which the periodic driving is switched on. Anomalous current is known to flow in junctions between superconductors when the junction region has spin orbit coupling and Zeeman field~\cite{yokoyama14,campagnano15,mintillo18}. In this work, we have shown that a nonequilibrium version of anomalous Josephson effect manifests due to a difference in phases of the driving potentials of Floquet superconductors and the effect survives in the limit when the driving is switched on adiabatically. By measuring the Josephson current in such driven systems, the effect can be observed in experiments. 
 
 A recent work studies a way of realizing qubits using Floquet Majorana fermions wherein adiabatically increasing the frequency results in stable qubits opposed to adiabatically increasing the amplitude of the driving~\cite{matthies22}. 
 Periodically driven systems connected to superconducting leads have been realized experimentally~\cite{park2022}.  Floquet topological insulators have been experimentally realized~\cite{gedik13} and it is possible to engineer superconductivity in topological insulators~\cite{hor10,sajadi18}. Superconductivity and magnetic field together in topological insulators are known to produce Majorana fermions. This suggests that application of proximity superconductivity and a magnetic field in Floquet topological insulator is a possible route to generate Floquet Majorana fermions. Floquet topological matter has been realized in acoustic~\cite{peng2016} and photonic~\cite{rechtsman2013,maczewsky2017} systems. Also, there has been a proposal for realizing Floquet topological superconductors using light~\cite{dehg21}. Many effects studied in this work also show up in non-topological Floquet superconductors, though much smaller in magnitude. Hence, realization of Floquet superconductor is an important step in experimental investigation of the effects predicted in this work. 
\acknowledgements
The author thanks Diptiman Sen,  Arijit Saha and Arijit Kundu for useful discussions.  The author acknowledges financial support by DST, India through DST-INSPIRE Faculty Award (Faculty Reg. No.~:~IFA17-PH190). 
\bibliography{reffajj}

\end{document}